\documentclass[aps,prb,floatfix,twocolumn,superscriptaddress,amssymb,amsmath]{revtex4-1}

\usepackage{graphicx}
\usepackage{dcolumn}
\usepackage{bm}
\usepackage{amsmath}
\usepackage{color}
\usepackage{float}
\usepackage{subfigure}
\usepackage{dsfont}
\usepackage{txfonts}
\usepackage{wasysym}
\usepackage{ifsym}

\newcommand{\veck}{\mathbf{k}}

\usepackage{multirow}

\makeatletter

\newcommand{\Rmnum}[1]{\expandafter\@slowromancap\romannumeral #1@}
\makeatother

\begin{document}

\title{Magnetic-field-modulated Kondo effect in a single-magnetic-ion
  molecule}

\author{Javier I. Romero}

\affiliation{Department of Physics, University of Central Florida,
  Orlando, Florida 32816-2385, USA}

\author{E.~Vernek}

\affiliation{Instituto de F\'isica, Universidade Federal de
  Uberl\^andia, Uberl\^andia, Minas Gerais 38400-902, Brazil}

\affiliation{Instituto de F\'{i}sica de S\~ao Carlos, Universidade de
  S\~ao Paulo, S\~ao Carlos, S\~ao Paulo 13560-970, Brazil}

\author{G.~B.~Martins} 

\affiliation{Department of Physics, Oakland University, Rochester, Michigan
  48309, USA}

\author{E.~R.~Mucciolo}

\affiliation{Department of Physics, University of Central Florida,
  Orlando, Florida 32816-2385, USA}

\keywords{Magnet molecule}

\begin{abstract}
We study numerically the low-temperature electronic transport
properties of a single-ion magnet with uniaxial and transverse spin
anisotropies. We find clear signatures of a Kondo effect caused by the
presence of a transverse (zero-field) anisotropy in the molecule.
This Kondo effect has an SU(2) pseudo-spin character, associated 
with a doublet ground state of the isolated molecule, which results from the
transverse anisotropy. Upon applying a transverse magnetic field to
the single-ion magnet, we observe oscillations of the Kondo effect due
to the presence of diabolical  points (degeneracies) of the energy
spectrum of the molecule caused by geometrical phase interference
effects, similar to those observed in the quantum tunneling of
multi-ion molecular nanomagnets. The field-induced lifting of the
ground state degeneracy competes with the interference modulation,
resulting in some cases in a suppression of the Kondo peak.
\end{abstract}

\date{\today}

\maketitle


\section{Introduction} 

In recent years, there has been a growing interest in the development
and study of single-ion magnets
(SIMs) \cite{Ishikawa2003,Ishikawa2007}. SIMs are synthesized with $3d$
transition metal or $4f$ rare-earth single-ion magnetic centers, thus
generating a simpler class of molecules whose properties are similar
to those of single-molecule magnets
(SMMs) \cite{Christou2000,Gatteschi2007,Wernsdorfer2008,Friedman2010}.
Among their characteristics, SIMs exhibit slow spin
relaxation \cite{Ishikawa2013,Zadrozny2011}, quantum tunneling of the
magnetization \cite{Ishikawa2005}, spin crossover regimes, and less
decoherence, thus making them suitable for qubit
applications \cite{Baldovi2012,Perez2012}. Moreover, one of the most
important characteristics of SIMs is that their magnetic anisotropies
can be tailored by chemical modification of their
ligands \cite{Gomez2013}. Recently, a ligand-modified SIM showed
remarkable robustness in its magnetic behavior in both solid state and
solution, with an observed magnetic hysteresis up to
4K \cite{DaCunha2013}. These properties motivate the study of
electronic transport in SIMs, given the importance that nanoscale
magnetic devices may have in future information technology
applications.

In this work we investigate the Kondo effect in a SIM in
the presence of spin anisotropies, and the effect of macroscopic spin
interference effects in the tunneling of the magnetization.
For these studies we employ the numerical renormalization-group method
(NRG) \cite{Bulla2008}, thus including the coupling of the
SIM to the itinerant electrons in the leads in a nonperturbative and
coherent way. We consider two different effective coordination
geometries of the SIM: one that induces only an uniaxial spin
anisotropy and another that induces both uniaxial and transverse spin
anisotropies. Upon the application of a transverse magnetic field
(perpendicular to the easy axis), the energy splitting
of the ground and first excited states is modulated (as a function of
the field magnitude) and oscillates periodically due to a Berry-phase
interference effect\cite{Garg1993}, i.e., the interference of spin
tunneling paths of opposite windings.

     Previous work by one of the authors studied Berry-phase oscillations
of the Kondo effect in SMMs using a poor man's scaling
approach \cite{Leuenberger2006}. In this context, the nature of the
effective coupling between the itinerant electrons and the molecule
magnetization was also evaluated \cite{Gonzalez2008}. Furthermore,
nonperturbative calculations have been done to simulate the many-body
charge transport properties of anisotropic magnetic impurities
adsorbed onto a Cu/Cu(100) surface \cite{Zitko2010}, as well as
different Kondo impurity Hamiltonians were used to model the static
and dynamical properties of magnetic atoms adsorbed in nonmagnetic
surfaces \cite{Zitko2008,Schnack2013}. The NRG method was previously
used to study the Kondo and Berry-phase interference effects in the
electronic transport through
SMMs\cite{Romeike2006a,Romeike2006b,Roosen,Misiorny} (see
Ref.~\onlinecite{Wegewijs2011} for a review). However, in those
studies the interaction between the itinerant electron spin and the
molecule magnetization was mainly phenomenological, namely, assumed
to take the isotropic form
\begin{equation}
H_{\textrm{Kondo}} = J_{\rm K} \vec{S}_{\textrm{SMM}} \cdot \vec{s},
\label{H_Kondo}
\end{equation}
where $\vec{s}$ is the itinerant electron-spin operator and
$\vec{S}_{\textrm{SMM}}$ is the total spin of the SMM. In this work we
propose a microscopic model that allows us to describe the Kondo
features of a SIM in terms of more fundamental parameters. For this
model, we find that, at zero magnetic field, an enhancement of the
conductance at low temperatures occurs whenever the SIM has a
coordination geometry that induces a transverse spin anisotropy for
both the ground and charged states of the molecule. The enhancement of
the conductance is caused by the occurrence of a pseudospin 1/2 Kondo
effect (originating from the transverse spin anisotropy), even though
the total spin of the SIM is larger than 1/2. Furthermore, we find
that the conductance through the SIM is modulated due to interference
effects in the spin tunneling paths of the SIM under an applied
transverse magnetic field.  As a result, the Kondo effect reemerges at
a nonzero magnetic-field value that is very accessible to experiments.

The paper is organized as follows: In Sec. \Rmnum{2} we
introduce the model Hamiltonian of the system. We also present the
energy spectrum and eigenstates of the isolated SIM in the
absence and in the presence of transverse anisotropy and magnetic-field terms, highlighting the periodic modulation of the spectrum by
the magnetic field. In Sec. \Rmnum{3} we evaluate the conductance of
the molecule as a function of transverse magnetic field. 
In Sec. \Rmnum{4} we summarize our conclusions.

\section{SIM model and Berry-phase modulation of the energy spectra} 

\begin{figure}
\includegraphics[width=.95\columnwidth]{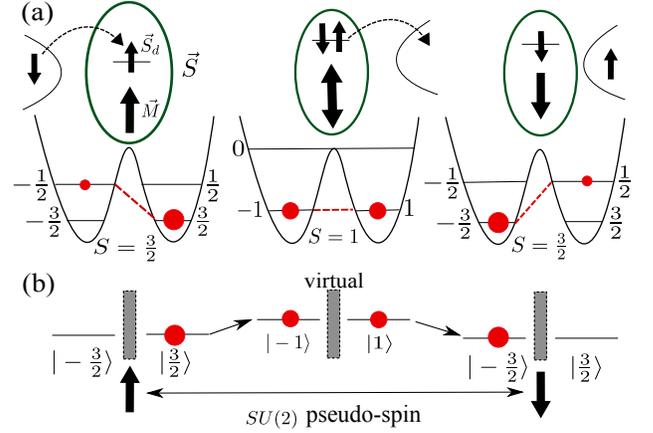}
\caption{\label{fig1} (Color online) (a) Schematic representation of
  the SIM with cotunneling processes that result in the pseudospin
  SU(2) Kondo effect: spin-down electron tunnels into predominantly
  $|3/2\rangle$ SIM ground state (left), resulting in a $|-1 \rangle +
  |1 \rangle$ virtual state (center). The subsequent tunneling out of
  a spin-up electron takes the SIM to a predominantly $|-3/2 \rangle$
  SIM ground state (degenerate with the initial $|3/2\rangle$
  state). A coherent sequence of these processes screens the
  pseudo-spin doublet ground state. (b) An equivalent schematic
  representation in the pseudo-spin picture, where the gray rectangle
  indicates the tunneling barrier between SIM states with opposite
  $S_z$ values.}
\end{figure}

We consider a transition metal SIM consisting of an $M=1$ magnetic moment
coupled to a $d$ orbital through a ferromagnetic Hund's interaction
[see Fig.~\ref{fig1}(a)]. The $d$ orbital has a spin $S_{d}=1/2$ when
singly occupied. In addition, the total magnetic moment of the
molecule, $\vec{S} = \vec{M} + \vec{S}_d$, has uniaxial and transverse
spin anisotropies induced by a ligand environment, with the easy axis
of the SIM along the $z$ direction. In the weak ligand-field regime,
the anisotropy parameters arise from a perturbation of the spin-orbit
coupling in the metal core using the orbital angular momentum states
of the free ion (no ligands) as the unperturbed eigenstates, thus the
orbital degrees of freedom are effectively integrated out in this
case. The molecule is connected to two leads of noninteracting
electrons only through the $d$ orbital [see Fig.~1(a)]. 
Here we assume that the electron transport occurs through
the SIM magnetic core, thus we do not consider transport through the
ligands. For simplicity, we also assume that the
anisotropy does not change with the charge state of the molecule. The
total Hamiltonian of the system is given by
\begin{equation}
H = H_{\textrm{M}} + H_{\textrm{leads}} + H_{\textrm{ML}},
\end{equation}
where
\begin{eqnarray}
\label{H_mol}
H_{\textrm{M}} & = & \sum_{\sigma} V_g n_{d\sigma} + \frac{U}{2}
N(N-1) - J_{\rm Hund} \vec{S}_{d} \cdot \vec{M} \nonumber \\ & &
-\ B_{1} S_{z}^{2} + B_{2} (S_{x}^{2} - S_{y}^{2}) - h_{x} S_{x},
\end{eqnarray}
\begin{eqnarray}
\label{H_lead}
H_{\textrm{leads}} = \sum_{\alpha, \veck,\sigma} \epsilon _{\alpha
  \veck\sigma}\, c^{\dagger}_{\alpha \veck\sigma} c_{\alpha
  \veck\sigma},
\end{eqnarray}
and 
\begin{equation}
\label{H_mol-lead}
H_{\textrm{ML}} = \sum_{\alpha,\veck,\sigma} \left( V_{\alpha \veck}\,
c^+_{ \alpha \veck\sigma } c_ { d\sigma } + \textrm{H.c.} \right).
\end{equation}
In Eqs.~(\ref{H_mol})--(\ref{H_mol-lead}), $V_g$ is the applied gate
voltage, $U$ and $J_{\rm Hund}$ are on-site direct and exchange
Coulomb interactions, respectively, with $n_{d\sigma}$ being the
number operator of the $d$ orbital and $N =
\sum_{\sigma}n_{d\sigma}$. We assume that the molecule is brought to
the particle-hole symmetric point through the gate voltage, therefore
$V_g = -U/2$. The positive coefficients $B_{1}$ and $B_{2}$
denote the easy-axis and transverse anisotropies of the components
$S_{j} = M_{j} + S_{dj}$ ($j=x,y,z$) of the total spin $S$ of the
molecule, and $h_{x} = g\mu_{B}H_{x}$ is a transverse magnetic field 
of amplitude $H_x$ applied along the $x$ axis.

Besides charge conservation in the molecule [U(1) symmetry], in the absence
of the transverse anisotropy and an in-plane external magnetic field, the $z$ 
component of the total spin of the molecule is a good quantum number, i.e., 
$[H_{\textrm{M}},S_{z}]=0$ and SU(2) spin-rotational symmetry is satisfied. In 
this case, states labeled by the $z$-components of the total spin are also eigenstates
of the Hamiltonian, and since $[S^2,S_z]=0$, the eigenstates will involve linear 
combinations $\vert S,S_{z}\rangle=\sum C[S_{dz},M_{z}; S,S_{z}] \vert S_{dz},M_{z}\rangle$, where the $C$'s
are Clebsch-Gordan coefficients. When $B_{2}>0$, rotational symmetry
is broken and the eigenstates of the Hamiltonian are written as
linear combinations of $\vert S,S_{z}\rangle$
states. Notice that even when the SU(2) rotational symmetry does not
hold anymore, $[H_{\textrm{M}},S^{2}]=0$, thus $S$ is always a
good quantum number.

$H_{\textrm{leads}}$ describes the leads, with $c^{\dagger}_{\alpha
  \veck\sigma}(c_{\alpha\veck\sigma})$ denoting creation
(annihilation) of an electron in lead $\alpha$ with energy
$\epsilon_{\alpha\veck\sigma}$, momentum $\veck$, and spin $\sigma =
\pm 1/2$. The tunneling term $H_{\textrm{ML}}$ contains the hopping
coupling $V_{\alpha\veck}$ between the $d$ orbital and the leads,
where $c_{d\sigma}$ ($c^{\dagger}_{\alpha\veck\sigma}$) annihilates
(creates) an electron with spin $\sigma$ in the SIM (lead).

\begin{figure}
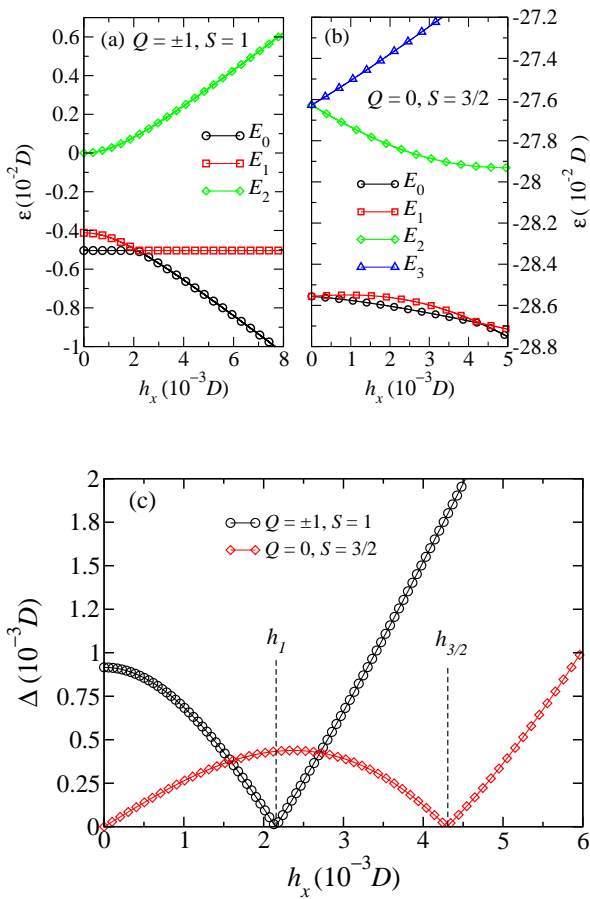

\centerline{
\includegraphics[width=0.9\columnwidth]{Splitting1and32NEWmod.eps}}
\vspace{0.8cm}
\centerline{
\includegraphics[width=0.9\columnwidth]{SplittingNEWmod.eps}}
\caption{\label{fig2} (Color online) Energy eigenvalues $E_i$ and
  splitting ($\Delta=E_1-E_0$) for the bare molecule (not connected to
  the leads), where $E_{i}$ is the $i$-th eigenvalue. The interaction
  parameters (in units of $D$) are: $V_g=-0.25$, $U=0.5$, $J_{\rm
    Hund}=5\times10^{-2}$, $B_{1}=4.6\times 10^{-3}$, and
  $B_{2}=4.6\times10^{-4}$. (a) Eigenvalues of the double occupied (or
  empty) $Q=\pm 1$, $S=1$ charge-spin sector versus $h_{x}/D$. (b)
  Eigenvalues of the neutral $Q=0$, $S=3/2$ charge-spin sector versus
  $h_{x}/D$. (c) Energy splitting $\Delta$ for the $Q=\pm1$, $S=1$
  [(black) circles] and $Q=0$, $S=3/2$ [(red) squares] charge-spin
  sectors.}
\end{figure}

Figures \ref{fig2}(a) and \ref{fig2}(b) show the magnetic-field effect
on the set of eigenvalues corresponding to the charge sectors $Q = \pm
1$ and $Q=0$, respectively (where $Q = N-1$). Fig. \ref{fig2}(c) shows the
tunneling splitting $\Delta = \vert E_{0}-E_{1}\vert$ between the two
lowest energy states corresponding to total spin sectors $S=1$ and
$S=3/2$. Note that these results were obtained for an isolated SIM,
i.e., not connected to the leads. For a molecule described by the
anisotropy terms in Eq. (\ref{H_mol}), the oscillations of the tunnel
splittings in Fig.~\ref{fig2}(c) result from a Berry-phase
interference effect \cite{Garg1993,Wernsdorfer2008}. At zero field,
spin parity dictates the nature of this interference \cite{Loss1992}:
the energy splitting is zero for a half-integer spin ($S=3/2$) (the
states form a Kramer's doublet) while it has a finite value for an
integer spin ($S=1$), as shown in Fig.~2(c). When a magnetic field
perpendicular to the easy axis is applied, time-reversal symmetry is
broken and oscillations of the splitting occur. Moreover, if the
direction of the magnetic field is along the hard-anisotropy axis
(making an angle of $0$ or $\pi$ with the $x$ axis), the interference
is maximum and the splitting goes to zero at the so-called diabolical
points $h_{1}$ and $h_{3/2}$ \cite{Berry1983}. At these special points,
seen also in the level crossings in Figs.~\ref{fig2}(a) and
\ref{fig2}(b), time-reversal symmetry of the corresponding spin sector
is effectively restored and a change in the symmetry of the ground and
first excited spin states occurs.

\begin{figure}
\centerline{
\includegraphics[width=1.0\columnwidth]{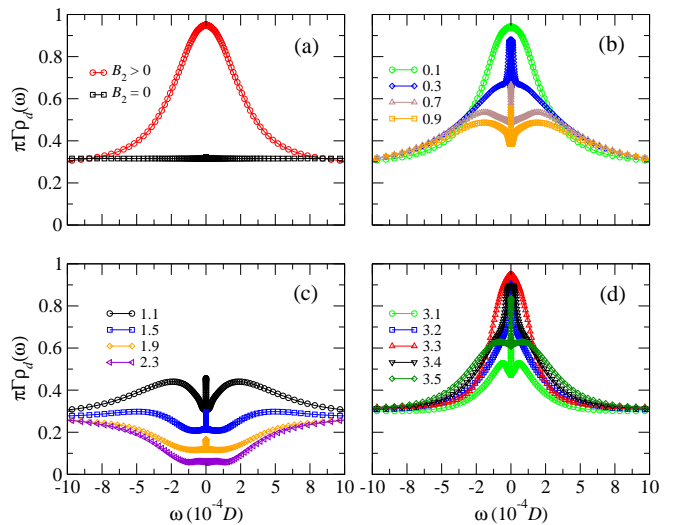}}
\caption{\label{fig3} (Color online) LDOS versus $\omega$ results for
  (a) $ h_{x}/D = 0$, $B_{2}/D=0$ [(black) squares] and
  $B_{2}/D=4.6\times10^{-4}$ [(red) circles]. In (b), (c), and (d)
  $0.1 \leq h_{x}/(10^{-3} \times D) \leq 3.5$ and
  $B_{2}/D=4.6\times10^{-4}$. The other interaction parameters are the
  same as in Fig.~\ref{fig2}, and $\Gamma = 9 \times 10^{-2}D$.}
\end{figure}

\section{Berry-phase modulation of electronic transport} 

We now allow the coupling of the SIM to the leads to be turned on and study the 
the low-temperature transport features of the SIM model by using the NRG
method \cite{Bulla2008}. We assume that the leads have a semi-circular
density of states (DOS) $\rho_{\textrm{c}}(\omega)=(1/2\pi
D^2)\sqrt{D^2-\omega^2}$, where $D$ is the half-width of the
conduction band, which is taken as our energy unit. Note that the use
of a flat DOS leads to essentially the same results. We consider the
tunneling coupling $V_{\veck,\alpha}=V$ to be independent of $\veck$
and equal for both leads, as well as set $k_{B}=1$. The hybridization
constant $\Gamma=\pi V^2\rho_{\textrm{c}}(0)$ is set to
$9\times10^{-2}\, D$. In our NRG calculations, we have 
used a discretization parameter $\Lambda = 2.5$ (keeping $3000$ 
states at each iteration for all calculations, except for the local density 
of states (LDOS) (see Fig.~3), where $\Lambda =2.0$ was used to smoothen 
out the oscillations around the Fermi energy. We verified that the fix 
point is reached before 80 sites of the Wilson's chain for both values of 
$\Lambda$, although we went up to $100$ sites for each calculation. 
We  also found that the phase shifts $\eta$ extracted from the numerical 
data were consistent with the existence of a Kondo effect ($\eta = \pi/2$). 

The dependence of the the $d$-level LDOS with the
applied transverse magnetic field around $\omega=0$ (i.e., near the Fermi
energy) is shown in Fig.~\ref{fig3}. At zero magnetic field [panel
  (a)], the Kondo resonance is absent for zero transverse anisotropy
($B_{2}=0$) since there is no tunneling between the $z$ components of
the {\it virtual} spin states $S_z=\pm 1$, thus the pseudo-spin
flipping of the initial state cannot take place. The finite
  LDOS for $B_2=0$ in Fig.~\ref{fig3}(a) arises because for $B_2= B_1 =
  h_x = 0$ the low-energy behavior of the systems corresponds to an
  underscreened $S = 3/2$ Kondo model, whose LDOS presents a cusp at
  the Fermi level \cite{Hewson2005}. By turning on the anisotropy term
  $B_1$ this Kondo effect can be partially suppressed (for
  sufficiently small values of $B_1$). This has been confirmed by
  doubling the value of $B_1$, in comparison to the one used in
  Fig.~\ref{fig3}(a), and noticing that the LDOS at the Fermi energy is 
further  suppressed (not shown).

When the transverse anisotropy is nonzero, the LDOS shows the
characteristic Kondo peak at the Fermi energy. In this case, one
expects an enhancement of the electronic conductance through the
molecule for energy scales smaller than $T_{K}$, where $T_K$ is the
Kondo temperature. The existence of a finite transverse anisotropy
($B_{2}>0$) in the molecule makes the {\it virtual} spin state of the
system a combination of $z$ components, i.e., $\vert\phi\rangle =
\frac{1}{\sqrt{2}} \left( \vert 1 \rangle - \vert -1 \rangle\right)$.
Thus, after an electron tunnels into the SMM, the molecule has a 50\%
probability of being in a spin $S_z=\pm 1$ state [center double well
  in Fig.~\ref{fig1}(a)], making the formation of a Kondo singlet
possible, through a sequence of cotunneling processes, as illustrated
in Fig.~\ref{fig1}.

As the magnetic field is turned on and increased, the initial doublet
ground state of the system becomes gapped in energy, and a suppression
of the Kondo peak takes place, as seen in Figs.~3(b) and
3(c). Interestingly, the Kondo peak eventually resurfaces again,
reaching a maximum height at about $h_{x}/D=3.3\times10^{-3}$ [see
  panel (d)]. This is because the energy gap created by the magnetic
field is driven to zero at the renormalized $\tilde{h}_{3/2}$
diabolical point.\cite{obs} Thus, Berry-phase interference is behind
the Kondo effect revival seen at this point.

\begin{figure}
\centerline{
\includegraphics[width=0.95\columnwidth]{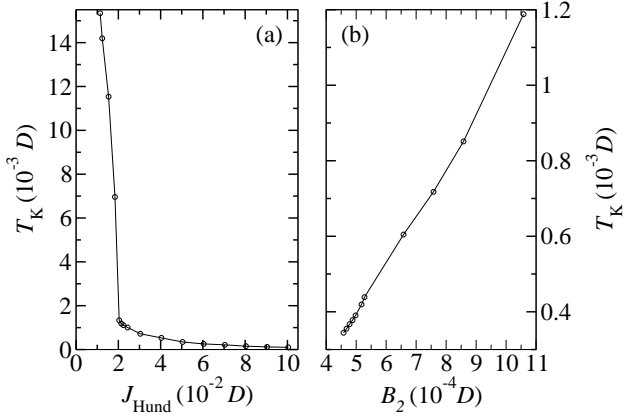}}
\caption{\label{fig4} Kondo temperature as a function of (a) the
  exchange coupling $J_{\rm Hund}$ (for $B_{2}/D=4.6\times10^{-4}$)
  and (b) the transverse anisotropy parameter $B_{2}$ (for $J_{\rm
    Hund}/D=5\times10^{-2}$). Other parameter values are
  $V_g/D=-0.25$, $U/D=0.5$, $B_{1}/D=4.6\times10^{-3}$, and
  $h_x=0$. $T_K$ is obtained from the LDOS data as the full width at
  half maximum of the $\omega=0$ peak.}
\end{figure}

When the in-plane anisotropy [the $B_2$ term in Eq.~(\ref{H_lead})] is
strong enough, the temperature threshold for the zero-field Kondo
singlet formation is $T_{K}/D \approx 5.1\times 10^{-4}$, which
corresponds to the $J_{\rm Hund} > T_{K}$ regime. In this regime,
conduction electrons see the whole spin $\vec{S}$ of the
molecule,\cite{Vernek2011} but only screen the doublet formed by
maximum spin components of $S$, namely, a pseudospin 1/2 local degree
of freedom, due to the in-plane anisotropy (see Fig.~\ref{fig1}). In
the opposite case, $J_{\rm Hund} < T_{K}$, the coupling between the
local moment $\vec{M}$ and the $d$ orbital is weak and only the spin
1/2 of the $d$ orbital is screened. However, this case is rather
difficult to occur in practice.

The Kondo temperature is quite sensitive to the exchange coupling
$J_{\rm Hund}$ and the transverse anisotropy parameter $B_{2}$. In
Fig.~\ref{fig4}(a) we show that $T_K$ increases considerably as $J_{\rm
  Hund}$ decreases below $2 \times 10^{-2} D$. This behavior
  is consistent with the fact that, as $J_{\rm Hund}$ approaches
  $T_{K}$, a transition in the screening regime takes place, resulting
  in the $d$ level being fully screened, thus causing the rapid
  increase of the Kondo temperature. On the other hand, as
long as $B_{2}$ is large enough and the condition $J_{\rm Hund} >
T_{K}$ is fulfilled, the screening of the pseudospin 1/2
dominates. Figure~\ref{fig4}(b) shows an increase of the Kondo temperature 
as $B_{2}$ becomes larger. This LDOS peak widening trend with increasing transverse
anisotropy is also experimentally meaningful since the molecular magnetic anisotropies 
can be tuned by ligand modification \cite{Gomez2013}, or by controlling 
the effective exchange coupling between the magnetic moment of the molecule 
and the electrodes \cite{Oberg2014},

\begin{figure}
\centerline{
\includegraphics[width=0.85\columnwidth]{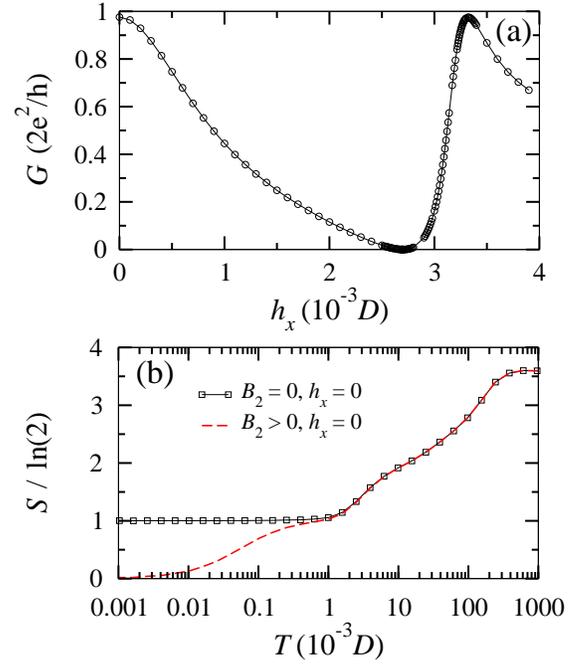}}
\caption{\label{fig5} (Color online) (a) Conductance through the
  molecule vs the transverse magnetic field $h_x$. (b) Entropy
  contribution of the SIM versus temperature.}
\end{figure}

The effect of the Berry-phase interference on the electronic transport
through the SIM is seen in Fig.~\ref{fig5}(a). The conductance is a
maximum for zero field and starts to decrease as the transverse
magnetic field is turned on, vanishing for $h_x/D \approx 2.7 \times
10^{-3}$. When the Berry-phase destructive interference sets in, the
energy splitting of the $Q=0$, $S=3/2$ doublet goes to zero and at the
$\Gamma$-renormalized diabolical point $\tilde{h}_{3/2}/D \approx 3.3
\times 10^{-3}$ the conductance reaches its maximum possible value
again. As one moves past this point, we observe that the conductance
decreases. We note that if one assumes that $D$ is likely to be of the
order of a few eV in experimental setups, the diabolical points can be
reached with relatively small fields (less than 1 T).

In Fig.~\ref{fig5}(b) we show the entropy contribution of the
molecule's magnetization. At high temperatures all spin states of the
molecule are available and we have ${\cal S} = \ln(2^{3.56})$, which
is consistent with the total number of accessible states $\Omega =
12$. As the temperature decreases we can approximately keep track of
the energy thresholds at which spin states of the molecule become
thermally forbidden. For $T/D < 6.16 \times 10^{-2}$, all high-energy
states ($Q = \pm 1$ sector) are suppressed. Thus, the only states
available are the ones in the $Q=0$ sector. At this temperature we can
have states belonging to different spin multiplets, i.e., $S=1/2$ or
$3/2$, with an energy splitting induced by the exchange interaction
$J_{\rm Hund}$. Singlet-triplet fluctuations are flushed away for $T/D
< 1.55 \times 10^{-2}$ (which is $\lesssim J_{\rm Hund}=5 \times
10^{-2}$). Moreover, as we cross into the $0< T/D < 8.14 \times
10^{-4}$ (which is $\lesssim B_1=4.6 \times 10^{-3}$) temperature
range, the double-degenerate first excited state with spin component
$S_{z}=\pm 1/2$ is not accessible anymore due to the uniaxial
anisotropy. Then, for these low temperatures, the molecule can only be
in the ground state $S_{z} = \pm 3/2$. Within this range, $B_{2}=0$
prevents the Kondo state, thus the entropy approaches $\ln{2}$. For
$T/D < 1.5 \times 10^{-4}$, which is below the Kondo temperature, the
entropy starts to approach zero since the pseudo-spin of the molecule
is locked into a Kondo singlet with the conduction electrons. Here
we also note that the temperature ranges where the Kondo effect sets
in are rather accessible to experiments: for $D$ of the order of a few
eV, $T_K$ is of the order of a few K.

\section{Conclusions} 

We have studied two interrelated effects in the electronic transport
through a SIM that arise from the properties of spin anisotropies and
spin-path interference behavior in this class of magnetic systems. The
first effect that we have found is that an enhancement of the
conductance of the molecule, namely, the Kondo effect, occurs whenever
there is a ligand distortion in the SIM that, in addition to creating
a strong easy-axis spin preferential direction, induces a transverse
spin anisotropy in the molecule. Moreover, we observe that the lifting
of the degeneracy due to the spin anisotropies transforms this problem
from an underscreened total spin $S=3/2$ onto a fully screened
pseudospin $S=1/2$ Kondo effect. The second effect is dependent on
the first condition, and consists of the modulation of the conductance
through the molecule upon the application of a transverse magnetic
field. This behavior is caused by a Berry-phase interference,
analogous to that observed in the quantum tunneling of the
magnetization of SMM crystalline systems. Both effects are rather
accessible to experiments and present an opportunity to probe the
interplay between electronic transport and magnetization tunneling
interference in molecular transistors.

\begin{acknowledgements}
The authors acknowledge financial support from
NSF ECCS-1001755 (E.R.M.); NSF DMR-1107994 and NSF MRI-0922811
(G.B.M.); CNPq, CAPES, and FAPEMIG (E.V.). J.R. acknowledges support
from the SBF-APS exchange program.
\end{acknowledgements}




\begin{thebibliography}{8}

\bibitem{Ishikawa2003} N. Ishikawa, M. Sugita, T. Ishikawa,
  S. Koshihara, and Y. Kaizu, J. Am. Chem. Soc. {\bf 125}, 8694
  (2003).

\bibitem{Ishikawa2007} N. Ishikawa, Polyhedron {\bf 26},
  2147Ð2153 (2007).

\bibitem{Christou2000} G. Christou, D. Gatteschi, D.~N. Hendrickson
  and R. Sessoli, MRS Bulletin {\bf 25}, 66 (2000).
  
\bibitem{Gatteschi2007} D. Gatteschi, R. Sessoli, and J. Villain, {\it
  Molecular Nanomagnets} (Oxford University Press, New York, 2007).

\bibitem{Wernsdorfer2008} W. Wernsdorfer, C. R. Chimie {\bf 11}, 1086
  (2008).

\bibitem{Friedman2010} J.~R. Friedman and M.~P. Sarachik,
  Annu. Rev. Condens. Matter Phys. {\bf 1}, 109 (2010).

\bibitem{Ishikawa2013} R. Ishikawa, R. Miyamoto, H. Nojiri,
  B.~K. Breedlove, and M. Yamashita, Inorg. Chem. {\bf 52}, 15 (2013).

\bibitem{Zadrozny2011} J.~M. Zadrozny and J.~R. Long, J.
  Am. Chem. Soc. {\bf 133}, 51 (2011).

\bibitem{Ishikawa2005} N. Ishikawa, J. Am. Chem. Soc. {\bf 127}, 3650
  (2005).

\bibitem{Baldovi2012} J. Baldov\'i, S. Cardona-Serra, J.  Clemente,
  E. Coronado, A. Gaita-Ari\~no, and A. Palii, Inorg. Chem. {\bf 51},
  22 (2012).

\bibitem{Perez2012} M. Mart\'inez-P\'erez {\it et al.},
  Phys. Rev. Lett. {\bf 108}, 247213 (2012).
  
\bibitem{Gomez2013} S. Gomez-Coca, E. Cremades, N.  Aliaga-Alcalde,
  and E. Ruiz, J. Am. Chem. Soc. {\bf 135}, 18 (2013).

\bibitem{DaCunha2013} T.~T. da Cunha {\it et al.},
  J. Am. Chem. Soc. {\bf 135}, 16332 (2013).
  
\bibitem{Bulla2008} R. Bulla, T.~A. Costi, and T. Pruschke,
  Rev. Mod. Phys. {\bf 80}, 395 (2008).
 
\bibitem{Garg1993} A. Garg, Europhys. Lett. {\bf 22}, 3 (1993).

\bibitem{Garg2003} A. Garg, E. Kochetov, K. Park, and M. Stone,
  J. Math. Phys. {\bf 44}, 48 (2003).

\bibitem{Leuenberger2006} M.~N. Leuenberger and E.~R. Mucciolo,
  Phys. Rev. Lett. {\bf 97}, 126601 (2006).

\bibitem{Gonzalez2008} G. Gonzalez, M.~N. Leuenberger, and
  E.~R. Mucciolo, Phys. Rev. B {\bf 78}, 054445 (2008).
  
\bibitem{Zitko2010} R. Zitko and T. Pruschke, New J. Phys. {\bf 12},
  063040 (2010).
  
\bibitem{Zitko2008} R. Zitko, R. Peters, and T. Pruschke, Phys. Rev. B
  {\bf 78}, 224404 (2008).

\bibitem{Schnack2013} M. H\"ock and J. Schnack Phys. Rev. B {\bf 87},
  184408 (2013).

\bibitem{Romeike2006a} C. Romeike, M.~R. Wegewijs, W. Hofstetter, and
  H. Schoeller, Phys. Rev. Lett. {\bf 96}, 196601 (2006).

\bibitem{Romeike2006b} C. Romeike, M.~R. Wegewijs, W. Hofstetter, and
  H. Schoeller, Phys. Rev. Lett. {\bf 97}, 206601 (2006).

\bibitem{Roosen} D. Roosen, M.~R. Wegewijs, and W. Hofstetter,
  Phys. Rev. Lett. {\bf 100}, 087201 (2008).

\bibitem{Misiorny} M. Misiorny, I. Weymann, and J. Barnas,
  Phys. Rev. B {\bf 86}, 035417 (2012).

\bibitem{Wegewijs2011} M.~R. Wegewijs, C. Romeike, and H. Schoeller,
  New J. Phys. {\bf 9}, 344 (2011).

\bibitem{Loss1992} D. Loss, D.~P. DiVincenzo, and G. Grinstein,
  Phys. Rev. Lett. {\bf 69}, 3232 (1992).
  
\bibitem{Hewson2005} W. Koller, A.~C. Hewson, and D. Meyer,
  Phys. Rev. B {\bf 72}, 045117 (2005)

\bibitem{Berry1983} M.~V. Berry and M. Wilkinson, Proc.  R. Soc. Lond
  A {\bf 392}, 15 (1984).

\bibitem{obs} Notice that the position of the diabolical points
  changes once the molecule is coupled to the leads.

\bibitem{Vernek2011} E. Vernek, F. Qu, F.~M. Souza, J.~C. Egues, and
  E.~V. Anda, Phys. Rev. B {\bf 83}, 205422 (2011).
  
\bibitem{Oberg2014}  J.~C. Oberg, M.~R. Calvo, F. Delgado, M. Moro-Lagares, D. Serrate, D. Jacob, J. Fernandez-Rossier, and C.~F. Hirjibehedin, 
Nat. Nanotechnol. {\bf 9}, 64 (2014).


\end{thebibliography}
\end{document}